%% file: s_vine_processes_20210702.tex
\documentclass[11pt]{article}

\input{preamble}   

\begin{document}
\title{Time series with infinite-order partial copula dependence}
 \author{Martin Bladt}
 \affil{Faculty of Business and Economics, University of Lausanne}
 \author{Alexander J.\ McNeil}
 \affil{The York Management School, University of York}
\date{2nd July 2020}
 \maketitle
 \begin{abstract}
 Stationary and ergodic time series can be constructed using an s-vine decomposition based on
sets of bivariate copula functions. The extension of such processes to infinite copula
sequences is considered and shown to yield a
rich class of models that generalizes Gaussian ARMA and ARFIMA processes to
allow both non-Gaussian marginal behaviour and a non-Gaussian
description of the serial partial dependence structure. Extensions of
classical causal and invertible representations of linear processes
to general s-vine processes are proposed and investigated. 
A practical and parsimonious method for parameterizing s-vine
processes using the Kendall partial autocorrelation function is
developed. The potential of the resulting models to give improved
statistical fits in many applications is indicated with an example
using macroeconomic data.
\end{abstract}
\section{Introduction}

The principal aim of this paper is to show that the s-vine (or
stationary d-vine) decomposition of a joint density provide a very natural vehicle for generalizing the class
of stationary Gaussian time series to permit both non-Gaussian marginal
behaviour and non-linear and non-Gaussian serial dependence
behaviour. In particular, this approach
provides a route to defining a rich class of tractable non-Gaussian ARMA and ARFIMA
processes; the resulting models have the potential to offer improved
statistical fits in any application where classical ARMA models or their
long-memory ARFIMA extensions are used.

Vine models of dependence have been developed in a series of publications
including~\cite{bib:joe-96,bib:joe-97},~\cite{bib:bedford-cooke-01,
  bib:bedford-cooke-01b,bib:bedford-cooke-02},~\cite{bib:kurowicka-cooke-06},~\cite{bib:aas-czado-frigessi-bakken-09}
and~\cite{bib:smith-min-almeida-czado-10}. There are a number of
different configurations for vines but the most suitable one for
longitudinal data applications is the d-vine, which is able
to describe strict stationarity of a random vector under some
additional translation-invariance restrictions on the vine
structure. In a recent
paper,~\cite{bib:nagler-kruger-min-20} investigate the vine 
structures that can be used to construct stationary multivariate time
series. Their results imply that, for univariate applications, the
d-vine is in fact the only structure for which translation-invariance
restrictions are sufficient to guarantee stationarity; we follow them
in referring to these restricted d-vines as stationary vines, or s-vines.

Vine models are best understood as copula models of dependence
 and there is now a large literature on copula models for time
series. While the main focus
of much of this literature has been on cross-sectional dependence between multiple time
series, there is also a growing literature on modelling serial
dependence within single series and lagged dependence across series.
The first-order Markov copula models investigated
by~\citet{bib:darsow-nguyen-olsen-92},~\citet{bib:chen-fan-06b},~\cite{bib:beare-10}
and~\cite{bib:domma-giordano-perri-09} are
simple examples of s-vine processes. A number of authors have written
on higher-order Markov extensions for univariate series or
multivariate series
including~\citet{bib:beare-seo-15}~\citet{bib:brechmann-christian-czado-15},
\citet{bib:louaiza-maya-et-al-18}
and~\citet{bib:nagler-kruger-min-20}. There is also literature
showing how these models may be adapted to the particular requirements
of time series showing stochastic volatility, including the
mixture-copula approach of~\citet{bib:louaiza-maya-et-al-18} and the
v-transform approach of~\citet{bib:mcneil-20} and~\citet{bib:bladt-mcneil-21}.

This paper makes the following novel
contributions to the development of time series models
based on vine copulas.
First, we suggest how s-vine models may be generalized to
infinite order and we propose accompanying generalizations of the classical concepts of
causality and invertibility for linear processes that may be applied
to s-vine processes. Second, we provide additional insight into the issues of
stability and ergodicity for s-vine processes and we show how finite or infinite copula sequences may be used to develop
non-linear filters of independent noise that generalize linear
filters. Finally, we propose a practical and parsimonious
approach to building s-vine processes in which copula
sequences are parameterized by a function that we call the
Kendall partial autocorrelation function; the latter may be borrowed
from other well-known processes, such as Gaussian ARMA or ARFIMA
processes, thus yielding natural non-Gaussian analogues of these models.

We believe that our approach may serve as a useful framework to faciliate further
study in the field. Several interesting theoretical
questions remain, particularly relating to necessary and sufficient
conditions for stability of models based on infinite copula sequences,
as well as the interplay of copula
sequences and long memory. However, on the practical side, the models
are already eminently usable; methods exist for estimation and random number generation, and
we further suggest methods for model validation using residuals. An example
shows the benefits that may arise from using these models.

The paper is structured as follows. Section~\ref{sec:s-vine-processes}
sets out notation and basic concepts and makes the connection between
s-vine copulas and s-vine processes; key objects in the development of
processes are sequences of functions that we refer to as Rosenblatt
functions. In Section~\ref{sec:finite-order}
we show that finite-order s-vine processes are Markov chains belonging
to the particular sub-category of non-linear state-space models. Section~\ref{sec:gaussian-processes} explains why
Gaussian processes form a sub-class of s-vine processes and shows how
classical theory for linear processes may be reinterpreted as a theory
of the behaviour of Rosenblatt functions. 
Section~\ref{sec:non-gauss-proc} uses the Gaussian analogy to suggest
requirements for stable, infinite-order, non-Gaussian s-vine
processes; a practical approach to model building is developed and
illustrated with an application to macroeconomic
data. Section~\ref{sec:conclusion} concludes. Proofs can be found in
Appendix~\ref{sec:proofs} while additional material on the Markov chain analysis of
finite-order processes is collected in Appendix~\ref{appendix:mark-chain-analys}.

\section{S-vine processes}\label{sec:s-vine-processes}

\subsection{S-vine copulas}

If a random vector $(X_1,\ldots,X_n)$ admits a
joint density $f(x_1,\ldots,x_n)$ then the latter may be decomposed as
a d-vine.
Writing $f_{X_i}$ for
the marginal density of $X_i$, the decomposition is
\begin{align}\label{D_vine_density}
f(x_1,\dots,x_n)&= \left(\prod_{i=1}^n
                  f_{X_i}(x_i)\right)
                  \prod_{k=1}^{n-1}\prod_{j=k+1}^{n}
                  c_{j-k,j|S_{j-k,j}}(F_{j-k|S_{j-k,j}}(x_{j-k}),F_{j|S_{j-k,j}}(x_{j}))
\end{align}
where $S_{j-k,j} =\{j-k+1,\ldots,j-1\}$ is the set of
indices of the variables which lie between $X_{j-k}$ and $X_{j}$,
$c_{j-k,j|S_{j-k,j}}$ is the density of the bivariate copula $C_{j-k,j|S_{j-k,j}}$ of the joint
distribution function (df) of $X_{j-k}$ and $X_{j}$ conditional on the
intermediate variables $X_{j-k+1},\ldots,X_{j-1}$, and
\begin{align}\label{pseudo_cond}
F_{i|S_{j-k,j}}(x)=\P(X_i \leq x|X_{j-k+1} = 
x_{j-k+1},\dots,X_{j-1} =
  x_{j-1}),\quad i \in \{j-k,j\}
\end{align}
denotes the conditional df of variable $i$ conditional on these
variables; note that $S_{j-1,j} =
\emptyset$ and so the conditioning set is dropped in this case. The decomposition~(\ref{D_vine_density}) implies a decomposition of the density $c(u_1,\ldots,u_n)$ of the
unique copula of  $(X_1,\ldots,X_n)$ which is given implicitly by
\begin{equation}
  \label{eq:16}
  c\left(F_1(x_1),\ldots,F_n(x_n)\right) =               \prod_{k=1}^{n-1}\prod_{j=k+1}^{n}
                  c_{j-k,j|S_{j-k,j}}(F_{j-k|S_{j-k,j}}(x_{j-k}),F_{j|S_{j-k,j}}(x_{j})).
\end{equation}

In practical modelling applications interest centres on models which
admit the \textit{simplified} d-vine
decomposition in which the copula densities $c_{j-k,j|S_{j-k,j}}$ do not
depend on the values of variables in the conditioning set $S_{j-k,j}$ and we can simply write
$c_{j-k,j}$; for more information about the
restriction, see~\cite{bib:haff-aas-frigessi-10}. Any set of copula densities
$\{c_{j-k,j} : 1 \leq k \leq n-1, k+1 \leq j \leq n\}$ and any set of
marginal densities $f_{X_i}$ may be used in the simplified version of~(\ref{D_vine_density}) to
create a valid $n$-dimensional joint density.

In this paper we are interested in strictly stationary stochastic processes
whose higher-dimensional marginal distributions are simplified
d-vines. As well as forcing $f_{X_1} = \cdots = f_{X_n}$, this requirement imposes translation-invariance conditions on the copula densities
$c_{j-k,j}$ and conditional dfs $F_{\cdot \mid S_{j-k,j}}$ appearing in the
    simplified form of~(\ref{D_vine_density}). It must
be the case that $c_{j-k,j}$ is the same for all $j\in\{k+1,\ldots,n\}$ and so each pair
copula density in the model can be associated with a lag $k$ and we can write
$c_k := c_{j-k,j}$ where $c_k$ is the density of some bivariate copula
$C_k$. The conditional dfs can be represented by two sets of functions
$\Rforward_k: (0,1)^k \times (0,1) \to (0,1)$ and $\Rbackward_k: (0,1)^k \times (0,1)
\to (0,1)$ which are defined in a recursive, interlacing fashion by
$\Rforward_1(u, x) =
h_1^{(1)}(u,x)$, $\Rbackward_1(u,x) = h_1^{(2)}(x,u)$ and, for $k \geq 2$,
\begin{equation}\label{eq:4b}
\begin{aligned}
   \Rforward_k(\bm{u}, x)  & =  h_k^{(1)}\left(  \Rbackward_{k-1}(\bm{u}_{-1}, u_1)  ,
                         \Rforward_{k-1}(\bm{u}_{-1}, x)   \right) \\
    \Rbackward_k(\bm{u}, x)  & =  h_k^{(2)}\left(  \Rbackward_{k-1}(\bm{u}_{-k}, x)  ,
                         \Rforward_{k-1}(\bm{u}_{-k}, u_k)   \right)
                            \end{aligned}
\end{equation}
where $h_k^{(i)}(u_1,u_2)  = \frac{\partial}{\partial u_i}C_k(u_1,u_2)$
and $\bm{u}_{-i}$ indicates the vector $\bm{u}$ with $i$th component
removed.

Using this new notation, we obtain a simplified form
of~(\ref{D_vine_density}) in which the density of the copula $c$ in~(\ref{eq:16}) takes the form
\begin{equation}
  \label{eq:4}
  c_{(n)}(u_1,\ldots,u_n)  = \prod_{k=1}^{n-1} \prod_{j=k+1}^n c_k \Big(\Rbackward_{k-1}(\bm{u}_{[j-k+1,j-1]},u_{j-k}),
    \Rforward_{k-1}(\bm{u}_{[j-k+1,j-1]},u_{j}) \Big) 
\end{equation}
where $\bm{u}_{[j-k+1,j-1]} = (u_{j-k+1},\ldots,u_{j-1})^\top$. Note
that,
for simplicity of formulas, we abuse notation by including terms involving
$\Rforward_0$ and $\Rbackward_0$; these terms should be interpreted as $\Rforward_0(\cdot,u) =
\Rbackward_0(\cdot,u) = u$ for all
$u$. Following~\cite{bib:nagler-kruger-min-20} we refer to a model
with copula density
of the form~(\ref{eq:4}) as an \textit{s-vine} or stationary d-vine.

If a random vector $(U_1,\ldots,U_n)$ follows the copula $C_{(n)}$
with density $c_{(n)}$ in~\eqref{eq:4} then
for any $k \in \{1,\ldots,n-1\}$
    and $j \in \{k+1,\ldots,n\}$,
    we have
    \begin{equation}\label{eq:7}
\begin{aligned}
  \Rforward_k(\bm{u}, x)  & =  \P(U_j \leq x \mid U_{j-k} = u_1, \ldots,
                         U_{j-1} = u_k) \\
    \Rbackward_k(\bm{u}, x)  & = \P(U_{j-k} \leq x \mid U_{j-k+1} = u_1, \ldots,
                         U_{j} = u_k) .
                       \end{aligned}
                       \end{equation}
and we refer to the conditional distribution functions $\Rforward_k$ and $\Rbackward_k$ as \textit{forward and backward
  Rosenblatt functions}. Henceforth we will often drop the
 superscript from the forward function and simply write $R_k =
 \Rforward_k$ to obtain less notationally cumbersome expressions.
The conditional densities corresponding to the Rosenblatt functions
may be derived from~\eqref{eq:4}. Writing $f_k$ for the density of the forward
Rosenblatt functions we obtain $f_1(u,x) = c_{(2)}(u,x) = c_1(u,x)$ and, for $k >1$
\begin{equation}\label{eq:1}
  f_k(\bm{u}, x) =
  \frac{c_{(k+1)}(u_{1},\ldots,u_{k},x)}{c_{(k)}(u_{1},\ldots,u_{k})}
  =
 \prod_{j=1}^kc_j \big(
  \Rbackward_{j-1}(\bm{u}_{[k-j+2,k]},u_{k-j+1}),
    \Rforwardb_{j-1}(\bm{u}_{[k-j+2,k]},x)
  \big).
\end{equation}

The following assumption will be in force throughout the remainder of
the paper.
\begin{assumption}\label{assumption:smooth-copulas}
All copulas $C_k$ used in the construction of s-vine models belong to
the class $\mathcal{C}^\infty$ of smooth functions with continuous partial derivatives of
all orders. Moreover their densities $c_k$ are strictly positive on $(0,1)^2$.
\end{assumption}
This assumption applies to all the standard pair copulas that are used
 in vine copula models (e.g.~Gauss, Clayton, Gumbel,
Frank, Joe and t), as well as non-exchangeable extensions following
the extension of~\cite{bib:liebscher-08} or mixtures of the kind
considered by~\cite{bib:louaiza-maya-et-al-18}. It ensures, among other
things, that for fixed
$\bm{u}$, the Rosenblatt functions are
bijections on $(0,1)$ with well-defined inverses. Let us write
$R^{-1}_k(\bm{u}, z)$ for the \textit{inverses of the Rosenblatt forward
functions}, satisfying  $R^{-1}_k(\bm{u}, z)= x$ if and only
if $\Rforwardb_k(\bm{u},x) = z$. Inverses can
 also be defined for the Rosenblatt
backward functions but will not be explicitly needed

In the sequel we refer to the copulas $C_k$ as \textit{partial} copulas. They should be distinguished from the bivariate \textit{marginal}
copulas given by $C^{(k)}(u,v) = \P(U_{j-k} \leq u , U_{j} \leq v) $
for any $j \in \{k+1,\ldots,n\}$. The two copulas are related by
the formula
\begin{align}
 C^{(k)}(v_1,v_2) &= \E \left(   \P(U_{j-k} \leq v_1,
                                    U_{j}\leq v_2 \mid U_{j-k+1},\ldots,U_{j-1})
                                    \right) \nonumber \\
  &= \E\left( C_k\left(
    \Rbackward_{k-1}((U_{j-k+1},\ldots,U_{j-1})^\top,v_1),
    \Rforwardb_{k-1}((U_{j-k+1},\ldots,U_{j-1})^\top,v_2) \right)\right)
    \nonumber \\
    &= \int_0^1 \cdots \int_0^1 C_k\left(
    \Rbackward_{k-1}(\bm{u},v_1),
    \Rforwardb_{k-1}(\bm{u},v_2) \right)
      c_{(k-1)}(\bm{u}) \rd u_1 \cdots \rd u_{k-1} .\label{eq:26}
\end{align}

\subsection{S-vine processes}

We will use the following general definition for an s-vine process. 

\begin{definition}[S-vine process]\label{def:d-vine}
  A strictly stationary time series $(X_t)_{t \in
      \Z}$ is an \textit{s-vine process} if for every $t \in \Z$ and $n\geq 2$ the
    $n$-dimensional marginal distribution of the vector $(X_{t},\ldots,X_{t+n-1})$ is
    absolutely continuous and admits a unique copula
    $C_{(n)}$ with a
joint density $c_{(n)}$ of the form~\eqref{eq:4}. An s-vine process $(U_t)_{t \in
      \Z}$ is an \textit{s-vine copula process} if its univariate
    marginal distribution
    is standard uniform.
  \end{definition}

  In the sequel our aim is to construct processes that conform to this
  definition and investigate their stability properties.
  Since s-vine processes can be endowed with any continuous univariate
  marginal distribution $f_X$, we will mostly investigate the
  properties of s-vine copula processes.

\subsection{A note on reversibility}

It is particularly common in applications of vine copulas to confine interest to
standard exchangeable copulas $C_k$. In this case the resulting s-vine
processes have the property of \textit{reversibility}. For any $\bm{u} =(u_1,\ldots,u_n)^\top \in (0,1)^n$ let us
write $\overline{\bm{u}} = (u_n, \ldots,u_1)^\top$ for the reversed
vector.
\begin{definition}
An s-vine copula process is reversible if for any $n \geq 2$ the
higher dimensional marginal copulas satisfy $C_{(n)}(\bm{u})  = C_{(n)}(\overline{\bm{u}})$.
  \end{definition}

  \noindent This is equivalent to saying that, for any $t,s \in \Z$ and
  any $n >2, $ the set of consecutive
  variables $(U_{t+1},\ldots,U_{t+n})$ from the process has the same
  distribution as the reversed vector
  $(U_{s+n},\ldots,U_{s+1})$. The
  process evolves forwards and backwards in a similar
  fashion, which may not be ideal for phenomena in which there is a clear
  temporal notion of causality; however, as soon as non-exchangeable copulas
  are included, the reversibility is broken.
  In summary we have
  following simple result.
  \begin{proposition}\label{prop:reversibility}
If a copula sequence $(C_k)_{k\in\N}$ consists of exchangeable copulas
then (i) the Rosenblatt forward and backward functions satisfy
$\Rbackward_k(\overline{\bm{u}}, x) = \Rforwardb_k(\bm{u},x)$ for all $(\bm{u},x) \in (0,1)^k
\times (0,1)$ and (ii) the resulting s-vine copula process is reversible.
    \end{proposition}

  \section{S-vine processes of finite order}\label{sec:finite-order}

  \subsection{Markov construction}
  The first class of processes we consider are s-vine copula processes of
  finite order $p$ which are constructed from a set of copulas
  $\{C_1,\ldots,C_p\}$ using the Markov approach described
  in~\citet[page 145]{bib:joe-15}. 
  Starting from a series of iid
uniform innovation variables 
$(Z_k)_{k \in \N}$ we can set $U_1 = Z_1$ and
\begin{equation}\label{eq:5}
  U_k =
R^{-1}_{k-1}\Big((U_{1},\ldots,U_{k-1})^\top,Z_k\Big),\quad k \geq 2.
\end{equation}
By using the inverses of the Rosenblatt
 forward functions we obtain, for any $n$, a random vector
  $(U_1,\ldots,U_n)$ which
  forms a finite realization from an s-vine process $(U_t)_{t \in \Z}$.
The copula $C_{(n)}$ of $(U_1,\ldots,U_n)$ has density $c_{(n)}$
  in~(\ref{eq:4}) but the copula densities $c_k$ appearing in
  this expression satisfy $c_k(u,v) = 1$ for $k >p$ and the s-vine is said
  to be truncated at order $p$. Moreover, since $h_k^{(1)}(u,v) = v$
  for $k > p$ it follows from~(\ref{eq:4b}) that $R_k(\bm{u}, x) =
  R_{k-1}(\bm{u}_{-1},x) = \cdots = R_p(\bm{u}_{[k-p+1,k]},x)$
  and the updating equation~\eqref{eq:5} satisfies
\begin{equation}\label{eq:55}
  U_k =
R^{-1}_{p}\Big((U_{k-p},\ldots,U_{k-1})^\top,Z_k\Big),\quad k
> p,
\end{equation}
showing the Markovian character of the finite-order process.

The recursive nature of the construction~\eqref{eq:5} means that there is an implied set of
functions that we will label $S_k: (0,1)^k \times (0,1) \to
(0,1)$ for $k \in\N$ such that
\begin{equation}
  \label{eq:15}
  U_k = S_{k-1}((Z_1,\ldots,Z_{k-1})^\top,Z_k)
  ,\quad k \geq 2.
\end{equation}
The functions $(S_k)_{k\in\N}$ satisfy $S_1(z_1,x) = R_1^{-1}(z_1,x)$
and
\begin{equation}
  \label{eq:23}
  S_k(\bm{z}, x) =
  R_k^{-1}\Big(\big(z_1,S_1(z_1,z_2),\ldots,S_{k-1}(\bm{z}_{[1,k-1]},z_k)\big),x\Big),\quad
   k \geq 2. 
\end{equation}
The identity~\eqref{eq:15} can be thought of as a
\textit{causal} representation of the process while the complementary identity $Z_k =
R_{k-1}((U_1,\ldots,U_{k-1})^\top,U_k)$ implied by~(\ref{eq:5}) can be
thought of as an
\textit{invertible} representation. We refer to the functions $(S_k)_{k\in\N}$ as
\textit{Rosenblatt inverse functions}; they should be distinguished
from the inverses of the Rosenblatt forward functions.


\subsection{Non-linear state space model}
   
      The s-vine process of order $p$ can be viewed as a
      $p$-dimensional Markov chain with state space $\mathcal{X} =
      (0,1)^p$. It is standard to treat Markov chains as being indexed
      by the natural numbers. To that end, for $t \in \N$, we introduce the vector-valued process $\bm{U}_t =
(U_{t},\ldots,U_{t+p-1})^\top$,  starting at $\bm{U}_1 = (U_1,\ldots,U_p)^\top$,
defined by the updating equation
$ \bm{U}_t = F(\bm{U}_{t-1},Z_t)$ where
\begin{equation}\label{eq:33}
F: (0,1)^p \times (0,1) \to (0,1)^p,\quad   F(\bm{u},z) = \Big(u_2,\ldots,u_p,R^{-1}_p(\bm{u}, z)\Big).
\end{equation}
The Markov chain described by~\eqref{eq:33}
 defines a \textit{non-linear state space (NSS) model} conforming exactly to
 the assumptions imposed
in~\citet{bib:meyn-tweedie-09} (see Section 2.2.2): under Assumption~\ref{assumption:smooth-copulas}
the updating function $F$ is a smooth ($\mathcal{C}^\infty$) function;
the state space  $\mathcal{X} = (0,1)^p$ is an open subset of
$\R^p$; the
uniform distribution of innovations $(Z_t)$ will be taken to be
supported on the open
set $(0,1)$.

Using standard arguments, the NSS model associated to \eqref{eq:33} can be shown to be a $\phi$-irreducible, aperiodic Harris
recurrent Markov chain and to
admit an invariant probability measure $\pi$ which is the
measure implied by the density $c_{(p)}$ given by~(\ref{eq:4}); we summarise
the arguments in Appendix~\ref{appendix:mark-chain-analys}.  This in
turn allows the ergodic theorem for Harris chains to be
applied~\citep[Theorem 13.3.3]{bib:meyn-tweedie-09} to conclude that
for any initial measure $\lambda$ the Markov transition kernel
$\mathsf{P}(\bm{x},\cdot)$ satisfies
\begin{displaymath}
  \left\lVert \int \lambda(\rd\bm{x}) \mathsf{P}^n(\bm{x},\cdot) - \pi(\cdot)\right\rVert
  \to 0,\quad n \to \infty
\end{displaymath}
where $\left\lVert \cdot \right\rVert$ denotes the total variation
norm. This is also sufficient for the strong law of large numbers (SLLN) to
hold~\citep[Theorem 17.0.1]{bib:meyn-tweedie-09}: for a function $g :\R^p \to \R$, if we define $S_n(g) =
\sum_{k=1}^n g(\bm{U}_k)$ and $\pi(g) = \int g(\bm{u}) c_{(p)}(\bm{u})
\rd \bm{u}$, then $\lim_{n\to\infty} n^{-1}S_n(g) = \pi(g)$, almost
surely, provided $\pi(|g|) < \infty$.

\begin{figure}[h]
\centering
\includegraphics[width=16cm,height=7cm]{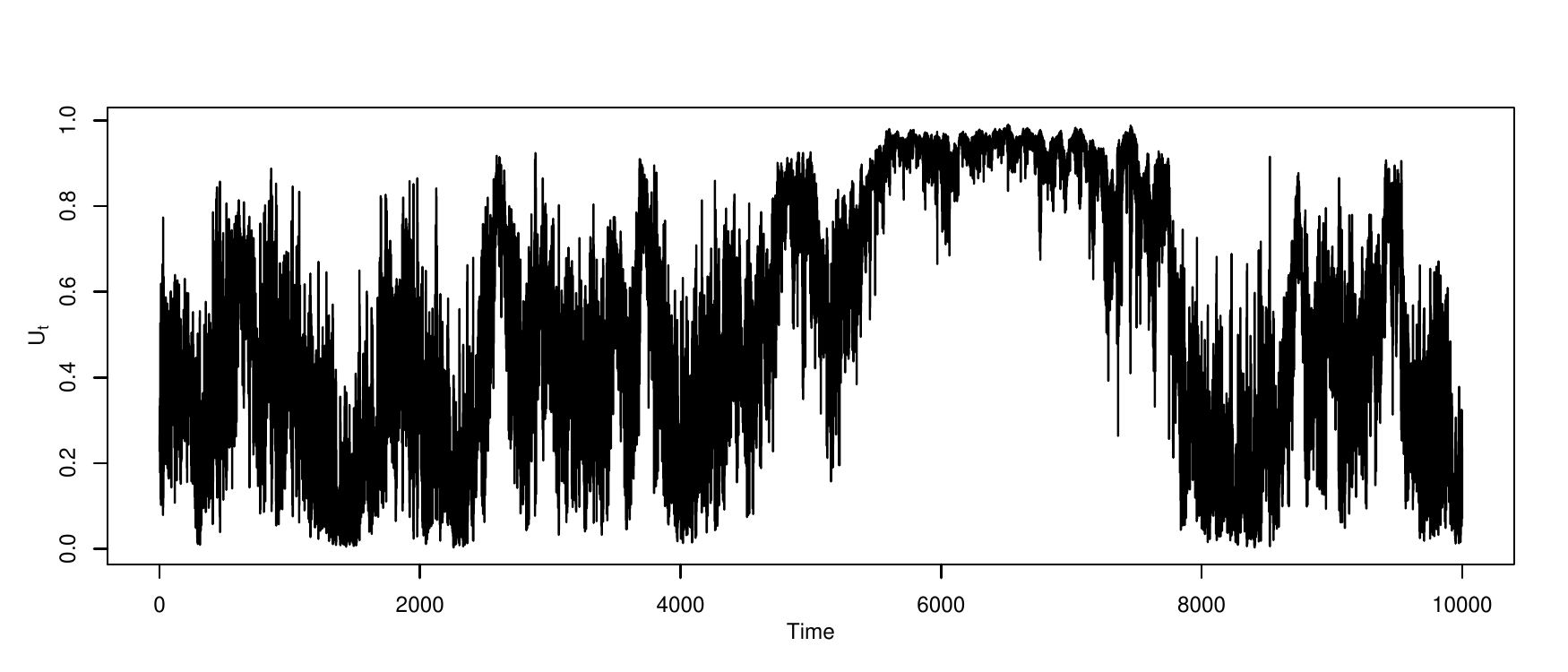}
\caption{Realisation of 10000 simulated
values from a process of order $k=3$ in which $C_1$ is a 180-degree
rotated
Clayton copula with parameter $\theta = 2$, $C_2$ is a Clayton copula
with $\theta=2$ and $C_2$ is a rotated Clayton  copula with
$\theta=4$.}
\label{fig:1}
\end{figure}

Although the Markov models are ergodic, they can exhibit
some very extreme
behaviour. Figure~\ref{fig:1} shows a realisation of 10000 simulated
values from a process of order $p=3$ in which $C_1$ is a 180-degree
rotated
Clayton copula with parameter $\theta = 2$, $C_2$ is a Clayton copula
with $\theta=2$ and $C_3$ is a rotated Clayton  copula with
$\theta=4$. There is a period of over 1500 successive values which are
all greater than 0.6. An observer of this
process who plots a histogram of the values in this period would have
difficulty
believing that the marginal distribution
is uniform.

There is a some literature on rates of mixing behaviour and 
rates of ergodic convergence for the case $p=1$, including~\cite{bib:chen-fan-06b},
\cite{bib:beare-10} and~\cite{bib:longla-peligrad-12}. The rates depend on the copulas that are chosen and,
in particular, their behaviour in joint tail regions. Tractible
general results on convergence rate for arbitrary sets of copulas in a model of arbitrary
order $p$ are not currently available.

\section{Gaussian processes}\label{sec:gaussian-processes}

 

Gaussian processes are processes whose finite-dimensional marginal
distributions are multivariate Gaussian. We will identify the term
Gaussian processes with \textit{non-singular} Gaussian processes
throughout; i.e.~we assume that the finite-dimensional marginal distributions of
Gaussian processes have invertible
covariance matrices and admit joint densities. Such processes
 represent a subclass of the s-vine processes.

\begin{proposition}\label{prop:gaussian-processes}
  \begin{enumerate}
  \item Every stationary Gaussian process is an s-vine process.
    \item Every s-vine process in which the pair copulas of
      the sequence $(C_k)_{k \in N}$ are Gaussian and the marginal
      distribution $F_X$ is Gaussian, is a Gaussian process.
  \end{enumerate}
\end{proposition}


\subsection{S-vine representations of Gaussian processes}
The first implication of Proposition~\ref{prop:gaussian-processes} is that every Gaussian process has a
unique s-vine-copula representation. This insight offers methods for
constructing or
simulating such processes as generic s-vine processes
using~(\ref{eq:5}) and estimating them using a
likelihood based on~(\ref{eq:4}).

Let $(X_t)_{t \in \N}$ be a stationary Gaussian process with mean $\mu_X$, variance $\sigma_X^2$ and
autocorrelation function (acf) $(\rho_k)_{k \in
\N}$; these three quantities uniquely determine a Gaussian
process. We assume the following:
\begin{assumption}\label{assumption:correlationtozero}
The acf $(\rho_k)_{k \in
\N}$ satisfies $\rho_k \to 0$ as $k \to \infty$.
\end{assumption}
\noindent It is well known that this is a necessary and sufficient
condition for a Gaussian process $(X_t)$ to be a mixing process and therefore ergodic~\citep{bib:maruyama-70,bib:cornfeld-fomin-sinai-82}. 

The acf
uniquely determines the partial autocorrelation function (pacf)
$(\alpha_k)_{k \in \N}$ through a one-to-one transformation~\citep{bib:barndorff-nielsen-schou-73,bib:ramsey-74}.
Since the partial autocorrelation of a Gaussian process is the
correlation of the conditional distribution of $(X_{t-k},X_{t})$ given the
intervening variables, the pair copulas in the s-vine copula representation are given by
$C_k = C^{\text{Ga}}_{\alpha_k}$.

 For $k \in \N$ let $\bm{\rho}_k=
 (\rho_1,\ldots,\rho_k)^\top$ and
let $P_k$ denote the correlation
matrix of $(X_1,\ldots,X_{k})$.  Clearly $P_1=1$ and, for $k >1$, $P_k$ is a symmetric Toeplitz
matrix whose diagonals are filled by the 
first $k-1$ elements of $\bm{\rho}_k$; moreover, $P_k$ is non-singular for
all $k$ under Assumption~\ref{assumption:correlationtozero}~\citep[Proposition 4]{bib:brockwell-davis-91}. The one-to-one series of
recursive transformations
relating $(\alpha_k)_{k \in \N}$ to $(\rho_k)_{k\in\N}$ is $ \alpha_1
= \rho_1$ and, for $k >1$,
\begin{equation}\label{eq:22}
  \begin{aligned}
     \alpha_k &= \frac{\rho_k -\bm{\rho}_{k-1}^\top
         P_{k-1}^{-1}\overline{\bm{\rho}}_{k-1 }}{ 1 - \bm{\rho}_{k-1}^\top
  P_{k-1}^{-1}\bm{\rho}_{k-1} } , \quad
\rho_k = \alpha_k\left(  1 - \bm{\rho}_{k-1}^\top
  P_{k-1}^{-1}\bm{\rho}_{k-1} \right) +  \bm{\rho}_{k-1}^\top
         P_{k-1}^{-1}\overline{\bm{\rho}}_{k-1}; 
     \end{aligned}
     \end{equation}
see, for example,~\citet{bib:joe-06} or
 the Durbin-Levinson
                                   Algorithm~\citep[Proposition
                                   5.2.1]{bib:brockwell-davis-91}.
                                   \begin{remark}
                                   Note that the restriction to
                                   non-singular Gaussian processes
                                   ensures that $|\rho_k| <1$ and
                                   $|\alpha_k| < 1$, for all $k \in
                                   \N$, and this is henceforth always assumed.
                                   \end{remark}

\noindent We review three examples of well-known Gaussian processes from
 the point of view of s-vine processes.
                                   \begin{example}[Gaussian ARMA models]
Any causal Gaussian ARMA($p$,$q$) model may be represented as an s-vine process
and full maximum likelihood estimation can be carried out using a
joint density based on~(\ref{eq:4}). If $\bm{\phi} =
(\phi_1,\ldots,\psi_p)^\top$ and  $\bm{\psi} =
    (\psi_1,\ldots,\psi_q)^\top$ denote the AR and MA parameters and
    $\rho_k(\bm{\phi},\bm{\psi} )$ the acf, then we can use the transformation~\eqref{eq:22}
 to parameterize~\eqref{eq:4} in terms of $\bm{\phi}$ and $\bm{\psi}$ using Gaussian pair
  copulas $C_k = C^{\text{Ga}}_{\alpha_k(\bm{\phi},\bm{\psi})}$.
In practice, this approach is more of theoretical
interest since standard estimation methods are generally much faster.
\end{example}

\begin{example}[Fractional Gaussian noise (FGN)]
  This process has acf given by
  \begin{displaymath}
    \rho_k(H) = \frac{1}{2}\left( (k+1)^{2H} + (k-1)^{2H}
      -2k^{2H}\right),\quad 0 < H <1,
  \end{displaymath}
  where $H$ is the Hurst exponent~\citep[see][for
  example]{bib:samorodnitsky-07}. Thus the transformation~\eqref{eq:22}
  may be used to parameterize~\eqref{eq:4} in terms of $H$ using Gaussian pair
  copulas $C_k = C^{\text{Ga}}_{\alpha_k(H)}$ and the FGN model may
  be fitted to data as an s-vine process and $H$ may be estimated.
\end{example}

\begin{example}[Gaussian ARFIMA models]\label{ex:arfima-processes}
The ARFIMA($p$,$d$,$q$) model with $ -1/2 < d < 1/2$ can be handled in a similar way to the
ARMA($p$,$q$) model, of which it is a generalization. In the case
where $p=q=0$,~\citet{bib:hosking-81} showed that
  \begin{equation}
    \label{eq:8}
      \alpha_k = \frac{d}{k-d},
      \quad k \in \N;
       \end{equation}
see also~\citet[Theorem 13.2.1]{bib:brockwell-davis-91}. The simple closed-form
expression for the pacf means that the  ARFIMA($0$,$d$,$0$) model is
even more convenient to treat as an s-vine than FGN; the two models
are in fact very similar in behaviour although
not identical. It is interesting to note that the pacf is not
summable and similar behaviour holds for some other ARFIMA
processes. For example, \cite{bib:inoue-02} has shown that for $p,q
\in \N\cup \{0\}$
and $0 < d < 1/2$ the pacf satisfies
$
  |\alpha_k| \sim d/k$ as $k \to \infty$.
\end{example}

\subsection{New Gaussian processes from s-vines}\label{sec:new-gauss-proc}

 A further implication of Proposition~\ref{prop:gaussian-processes} is that it shows how we can
 create and estimate some new stationary and ergodic Gaussian
 processes without setting them up in the classical way using
 recurrence equations, lag operators and Gaussian innovations. Instead
 we choose sequences of Gaussian pair copulas $(C_k)$ parameterized by
 sequencies of partial correlations $(\alpha_k)$.

 As in the previous section, we can begin with a
 parametric form
 for the acf $\rho_k(\bm{\theta})$ such that $\rho_k(
 \bm{\theta}) \to 0$ as $k \to \infty$ and build the model
 using pair copulas parameterized by the parameters $\bm{\theta}$ of the implied pacf
 $\alpha_k(\bm{\theta})$. 
 Alternatively we can choose a parametric form
 for the pacf $\alpha_k(\bm{\theta})$ directly.

 Any finite set
 of values $\{\alpha_1,\ldots,\alpha_p\}$
 yields an AR(p) model which is a special case
 of the finite-order s-vine models of
 Section~\ref{sec:finite-order}. However, infinite-order processes
 that satisfy Assumption~\ref{assumption:correlationtozero} are more
 delicate to specify. A necessary condition is that the
 sequence $(\alpha_k)$  satisfies $\alpha_k \to
 0$ as $k \to 0$, but this is not sufficient. To see this, note that if $\alpha_k = (k+1)^{-1}$, the
 relationship~(\ref{eq:22}) implies that $\rho_k = 0.5$ for all $k$
 which violates Assumption~\ref{assumption:correlationtozero}. A sufficient condition
 follows from a result of~\citet{bib:debowski-07}, although, in view
 of Example~\ref{ex:arfima-processes}, it is not a necessary
 condition:
\begin{assumption}\label{assumption:partialcorrelationsum}
The partial acf $(\alpha_k)_{k \in
\N}$ satisfies $\sum_{k=1}^\infty |\alpha_k| < \infty$.
\end{assumption}
 
\noindent\citet{bib:debowski-07} showed that,
if Assumption~\ref{assumption:partialcorrelationsum} holds, then the equality
\begin{equation}\label{eq:3}
  1 + 2\sum_{k=1}^\infty \rho_k = \prod_{k=1}^\infty \frac{1 +
    \alpha_k}{1 - \alpha_k}.
\end{equation}
also holds.
The rhs of~\eqref{eq:3} is a convergent product since absolute summability
ensures that the sums $\sum_{k=1}^\infty\ln (1
\pm \alpha_k)$ converge. This implies the convergence of
$\sum_{k=1}^\infty \rho_k$ which implies $\rho_k \to 0$ which in turn implies that
Assumption~\ref{assumption:correlationtozero} also holds, as we require.
  
 Assumption~\ref{assumption:partialcorrelationsum} still allows some quite
 pathological processes, as noted by~\citet{bib:debowski-07}. For
 example, even for a finite-order AR($p$) process
 with $\alpha_k \geq a > 0$ for $k \in \{1,\ldots,p\}$ and $\alpha_k =
 0$ for $k >p$ it follows that $\sum_{k=1}^\infty \rho_k \geq
 0.5(((1+a)/(1-a))^p -1)$ and this grows exponentially with $p$ leading
 to an exceptionally slow decay of the acf.
 
 \subsection{Rosenblatt functions for Gaussian processes}

For Gaussian processes the Rosenblatt functions and inverse Rosenblatt
functions take relatively tractable forms.

 \begin{proposition}\label{prop:gaussian-case}
Let $(C_k)_{k\in\N}$ be a sequence of Gaussian pair copulas with
parameters $(\alpha_k)_{k  \in \N}$ and assume that
Assumption~\ref{assumption:correlationtozero} holds. The forward Rosenblatt functions are given by
\begin{equation}\label{eq:9}
  R_k(\bm{u}, x ) = \Phi \left( \frac{\Phi^{-1}(x) - \sum_{j=1}^k
                        \phi^{(k)}_{j}
                        \Phi^{-1}(u_{k+1-j})}{\sigma_k}
                        \right), 
                      \end{equation}
                      where $\sigma_k^2 =\prod_{j=1}^i
                      (1-\alpha_j^2)$ and the coefficients
                      $\phi_j^{(k)}$ are given recursively by
                      \begin{equation}
                        \label{eq:11}
                        \phi_j^{(k)} =\begin{cases}
                          \phi_j^{(k-1)} -
                      \alpha_k\phi_{k-j}^{(k-1)},& j \in
                      \{1,\ldots,k-1\}, \\
                    \alpha_k,& j= k.\end{cases}
                    \end{equation}
                    The inverse Rosenblatt functions 
                    are given by
                    \begin{equation}
                      \label{eq:17}
              S_k(\bm{z}, x ) = \Phi \left( \sigma_k \Phi^{-1}(x) + \sum_{j=1}^k
                        \psi^{(k)}_{j}
                        \Phi^{-1}(z_{k+1-j})
                        \right),         
                      \end{equation}
                      where the coefficients $\psi_j^{(k)}$ are given
                      recursively by 
                      \begin{equation}
                        \label{eq:17b}
                \psi_j^{(k)} =  \sum_{i=1}^j \phi_i^{(k)}
                          \psi^{(k-i)}_{j-i} , \quad j \in
                          \{1,\ldots,k\},
                      \end{equation}
  where $\psi_0^{(k)} = \sigma_k$ for $k\geq 1$ and $\psi^{(0)}_0 =1$.
                    \end{proposition}

\noindent We can analyse the behaviour of the Rosenblatt and inverse Rosenblatt functions
as $k \to \infty$ in a number of different cases.

\paragraph{Gaussian processes of finite order.}
In the case of a Gaussian s-vine process of finite order $p$ we have, for $k >p$, that $\alpha_k = 0$, $\sigma_k = \sigma_p$ and $\phi_j^{(k)} = \phi_j^{(p)}$. If $(U_k)_{k \in \N}$ is constructed from $(Z_k)_{k\in\N}$ using
the algorithm described by~(\ref{eq:5}), and if we make the
substitutions $X_k = \Phi^{-1}(U_k)$ and $\epsilon_k = \Phi^{-1}(Z_k)$
as in the proof of Proposition~\ref{prop:gaussian-case}, then we
have that $X_k = \sum_{j=1}^{p} \phi_j^{(p)}X_{k-j} +
                         \sigma_{p} \epsilon_k$ for $k >p$, which is
                         the classical recurrence equation that defines
                         a Gaussian AR($p$) process; we also have that
                         $X_k = \sum_{j=1}^{k-1}
                         \psi_j^{(k-1)}\epsilon_{k-j} + \sigma_p
                         \epsilon_k$ for $k>p$. These two representations can be
                         written in invertible and causal forms as
                         \begin{equation}
                           \label{eq:25}
                           \epsilon_{k} = \sum_{j=0}^{p}
                        \tilde{\phi}^{(p)}_{j}
                        X_{k-j}        \quad\text{and}\quad                 
                                 X_{k} = \sum_{j=0}^{k-1}
                        \psi^{(k-1)}_{j}
                        \epsilon_{k-j}, \quad k >p,
                      \end{equation}
                        where $\tilde{\phi}^{(p)}_0 = 1/\sigma_p$,
    $\tilde{\phi}^{(p)}_j = - \phi^{(p)}_j/\sigma_p$ for $j >1$ and
    $\psi_0^{(k-1)} =\sigma_p$.

    Classical time series theory is concerned with
    conditions on the AR coefficients $\tilde{\phi}_j^{(p)}$ that allow us to
    pass to an infinite-order moving-average representation in the
    second series in~\eqref{eq:25}. In fact, by setting up our
    Gaussian models
    using partial autocorrelations,
    causality in the classical sense is guaranteed; this follows as a
    special case of
    Theorem~\ref{theorem:causal-invertible} below.

\paragraph{Gaussian processes with absolutely summable partial autocorrelations.}
We next consider a more general case where the process may be of
infinite order, but Assumption~\ref{assumption:partialcorrelationsum}
holds. To consider infinite-order models we now consider a process $(U_t)_{t\in\Z}$ defined on the
    integers. The result that follows
    is effectively a restating of a result
 by~\cite{bib:debowski-07} in the particular context of Gaussian s-vine copula processes.

 \begin{theorem}\label{theorem:causal-invertible}
Let $(U_t)_{t\in\Z}$ be a Gaussian s-vine copula process for which
the parameters $(\alpha_k)_{k\in\N}$ of the Gaussian pair copula
sequence $(C_k)_{k\in\N}$ satisfy
Assumption~\ref{assumption:partialcorrelationsum}. Then, for all $t$,
we have the almost sure limiting representations
\begin{align}
  U_t &=
        \lim_{k\to\infty}S_k((Z_{t-k},\ldots,Z_{t-1})^\top,Z_t)  \label{eq:17}
  \\
  Z_t &= \lim_{k\to\infty}R_k((U_{t-k},\ldots,U_{t-1})^\top,U_t)  \label{eq:18}
\end{align}
for an iid uniform innovation process $(Z_t)_{t\in\Z}$.
   \end{theorem}

\paragraph{Long-memory ARFIMA processes.}
As noted earlier, \cite{bib:inoue-02} has shown that the pacf of an
ARFIMA($p$,$d$,$q$) model with $0 <d < 0.5$ is not absolutely summable
and so Theorem~\ref{theorem:causal-invertible} does not apply in this
case. Nevertheless,~\citet[Section
13.2]{bib:brockwell-davis-91} show that the Gaussian process has a casual
representation of the form $   X_{t} = \sum_{j=0}^{\infty}
                        \psi_{j}\epsilon_{t-j}$
where convergence is now in mean square and the coefficients
are square summable, i.e.~$\sum_{j=0}^\infty \psi_j^2 < \infty$. Since
convergence in mean square implies convergence in probability, the continuous mapping theorem implies that a representation of the
form $ U_t =\lim_{k\to\infty}S_k((Z_{t-k},\ldots,Z_{t-1})^\top,Z_t) $
at least holds under convergence in probability.

\paragraph{A non-causal and non-invertible case.}\label{example:nonconvergent}
If $\alpha_k = 1/(k+1)$ for all $k$, then $\rho_k = 0.5$ and
both Assumptions~\ref{assumption:correlationtozero}
and~\ref{assumption:partialcorrelationsum} are violated.
It can be verified (for example by induction) that the recursive
formulas~(\ref{eq:11}) and~(\ref{eq:17b}) imply that
$\phi_j^{(k)} = 1/(k+1)$ and $\psi_j^{(k)} = \sigma_{k-j}/(k+2-j)$
for $j\geq 1$ (recall that $\psi_0^{(k)} = \sigma_k$). These
coefficient sequences are unusual; the coefficients
$\phi_j^{(k)}$ of the Rosenblatt function in~(\ref{eq:9}) place equal
weight on all past values $X_{k+1-j} = \Phi^{-1}(U_{k+1-j})$ while the coefficients
  $\psi_j^{(k)}$ of the inverse Rosenblatt function on the innovations
  in~(\ref{eq:17}) place weight $\psi_k^{(k)} = 1/2$ on the first value
  $\epsilon_1 = \Phi^{-1}(Z_{1})$ and decreasing weights on more recent
    values $\epsilon_j$, $j >1$.

    As $k \to \infty$, we do have
    $\sigma_k^2 = \prod_{j=1}^k (1 - 1/(k+1)^2) \to 1/2$, but, for
    fixed $j \geq 1$, the terms $\psi_j^{(k)}$ and $\psi_j^{(k)}$ both
    converge to the trivial limiting value 0 and we do not obtain
    convergent limiting representations of the form~(\ref{eq:17})
    and~(\ref{eq:18}) for $Z_t$ or $U_t$ in terms of past process values or past
    innovation values.

  \section{General s-vine processes}\label{sec:non-gauss-proc}

  We now consider infinite-order s-vine copula processes constructed from
  general sequences $(C_k)_{k \in \N}$ of pair copulas.

  \subsection{Causality and invertibility}

   The key consideration
  for stability of an infinite-order process is whether it admits a
  convergent causal representation. A process $(U_t)_{t\in\Z}$ with such a
  representation is a convergent non-linear filter of independent
  noise. It will have the property that $U_t$ and $U_{t-k}$ are
  independent in the limit as $k \to \infty$, implying mixing
  behaviour and ergodicity. We suggest the following
  definition of the causality and invertibility properties for a
  general s-vine process.

  \begin{definition}
    Let $(C_k)_{k\in\N}$ be a sequence of pair copulas and let
    $(R_k)_{k\in\N}$ and  $(S_k)_{k\in\N}$ be the corresponding
    Rosenblatt forward functions and Rosenblatt inverse functions defined by~(\ref{eq:4b}) and~(\ref{eq:23}).
    An s-vine copula process $(U_t)_{t\in\Z}$ associated with the
    sequence $(C_k)_{k\in\N}$ is strongly causal if
    there exists a process of iid uniform random variables
    $(Z_t)_{t\in\Z}$ such that~(\ref{eq:17}) holds almost surely for
    all $t$ and it is strongly
  invertible if representation~(\ref{eq:18}) holds almost surely
 for all $t$. If convergence in~(\ref{eq:17}) and~(\ref{eq:18}) only
 holds in probability, the process is weakly causal or weakly invertible.
\end{definition}

We know that Gaussian ARMA processes defined as s-vine
processes are always strongly causal (and invertible) and that the long-memory
ARFIMA($p$,$d$,$q$) process with $0 <d < 0.5$ is weakly causal.
When we consider sequences of Rosenblatt functions for sequences of
non-Gaussian pair copulas, proving causality appears to be
more challenging mathematically, since it is no longer a question of analysing the
convergence of series. In the next section we use simulations
to conjecture that causality holds for a class of processes defined
via the Kendall correlations of the copula sequence.

In a
  finite-order process the copula sequence for any lag $k$ greater
  than the order $p$ consists of independence copulas; it seems
  intuitively clear that, to obtain an infinite-order
  process with a convergent causal representation, the
  partial copula sequence $(C_k)_{k\in\N}$ should converge to the
  independence copula $C^\perp$ as $k \to \infty$. However, in view of Example~\ref{example:nonconvergent}, this is not a sufficient
condition and the speed of convergence of the copula sequence
is also important. Ideally we require conditions on the speed
of convergence
$C_k \to C^\perp$ so that the marginal copula $C^{(k)}$
in~(\ref{eq:26}) also tends to
$C^\perp$; in that case the variables $U_t$ and $U_{t-k}$ are
asymptotically independent as $k \to \infty$ and mixing
behaviour follows.

\subsection{A practical approach to non-Gaussian s-vines}\label{sec:pract-appr-non}

Suppose we take a sequence of pair copulas $(C_k)_{k\in\N}$ from some
parametric family and parameterize them in such a way that (i) the
copulas converge uniformly to the independence copula as $k \to
\infty$ and (ii) the \textit{level of dependence} of each copula $C_k$
is identical to that of a Gaussian pair copula sequence that gives
rise to an ergodic Gaussian process. The intuition here is that by
sticking close to the pattern of decay of dependence in a well-behaved
Gaussian process, we might hope to construct a stable causal process
that is both mixing and ergodic.

A natural way of making `level of dependence' concrete is to
consider the Kendall rank correlation function of the copula sequence,
defined in the following way.

    \begin{definition}
The Kendall partial autocorrelation function (kpacf)
$(\tau_k)_{k\in\N}$ asssociated with a copula sequence $(C_k)_{k\in\N}$ is given by
   \begin{displaymath}
     \tau_k = \tau(C_k), \; k \in \N,
   \end{displaymath}
   where $\tau(C)$ denotes the Kendall's tau coefficient for a copula $C$.
 \end{definition}

 For a Gaussian copula sequence with $C_k = C^{\text{Ga}}_{\alpha_k}$
 we have
\begin{equation}\label{eq:24}
  \tau_k = \frac{2}{\pi}\arcsin(\alpha_k).
\end{equation}
  As in
 Section~\ref{sec:new-gauss-proc}, suppose that
 $(\alpha_k(\bm{\theta}))_{k\in\N}$ is the pacf of a stationary and
 ergodic model Gaussian process parametrized by the parameters
 $\bm{\theta}$, such as an ARMA or ARFIMA model; this implies a
 parametric form for the kpacf $(\tau_k(\bm{\theta}))_{k \in \N}$. The
 idea is to choose a sequence of non-Gaussian pair copulas that shares
 this kpacf.

 A practical problem that may arise is that
 $\tau_k = \tau_k(\bm{\theta})$ can in theory take any value in $(-1,1)$; only
 certain copula families, such as Gauss and Frank, are said to be
 \textit{comprehensive} and yield any value for $\tau_k$. If we wish to use, for example, a sequence of Gumbel copulas to
 build our model then we need to find a solution for negative values
 of Kendall's tau. One possibility is to allow 90 or 270 degree
 rotations of the copula at negative values of $\tau_k$ and another is
 to substitute a comprehensive copula at any position $k$ in the
 sequence where $\tau_k$ is negative.

\begin{remark}
  Note that the assumption that the pair copulas $C_k$ converge to the
  independence copula has implications for
  using $t$ copulas $C^t_{\nu,\alpha}$ in this approach.
  The terms of the copula sequence $C_k = C^t_{\nu_k,\alpha_k}$ would
  have to satisfy $\nu_k \to \infty$ and $\alpha_k \to 0$ as $k \to
  \infty$; a model constructed from the sequence $C_k =
  C^t_{\nu,\alpha_k}$ for fixed $\nu$ would give rise to an
  irregular process. While
  the sequence $(\alpha_k)_{k\in\N}$ can be connected to the kpacf by the same
  formula~(\ref{eq:24}), the sequence $(\nu_k)_{k \in \N}$ is not fixed by the
  kpacf. It is simpler in this approach to work with copula families
  with a single parameter so that there is a
  one-to-one relationship between Kendall's tau and the copula parameter.
\end{remark}

To compare the speed of convergence of the copula filter for different copula sequences
sharing the same kpacf, we conduct some simulation experiments.
For fixed $n$ and for a fixed realization $z_1,\ldots,z_n$ of
independent uniform noise we plot the points $(k,
S_k(\bm{z}_{[n-k,n-1]},z_n))$ for $k \in
\{1,\ldots,n-1\}$. We expect the points to converge to a fixed value
as $k \to n-1$, provided we take a sufficiently large value of $n$. When the copula sequence consists of Clayton
copulas we will refer to the model as a Clayton copula filter;
similarly Gumbel copulas yield a Gumbel copula filter; and so on. The
following examples suggest that there are some differences in the
convergence rates of the copula filters; in particular, the filters
based on sequences of tail-dependent copulas like Clayton, Joe and
Gumbel show slower convergence. 

\begin{figure}[h]
\centering
\includegraphics[width=\textwidth]{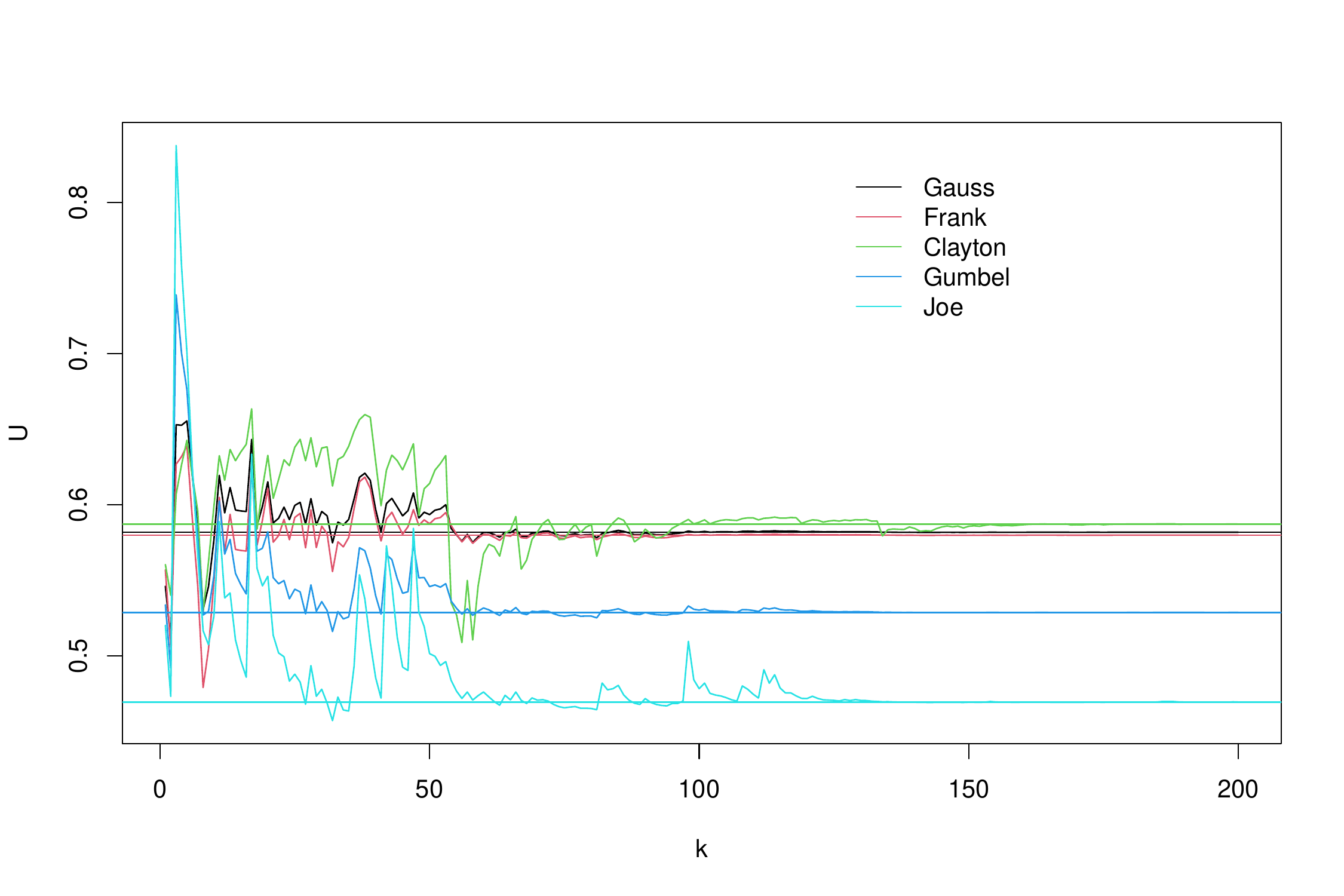}
\caption{Plots of $(k,
S_k(\bm{z}_{[n-k,n-1]},z_n))$ for $k \in
\{1,\ldots,n-1\}$ for the copula filters of ARMA(1,1)
models; see Example~\ref{example:arma}. Horizontal lines show ultimate values $S_{n-1}(\bm{z}_{[1,n-1]},z_n)$.}
\label{fig:2}
\end{figure}

\begin{example}[Non-Gaussian ARMA(1,1) models]\label{example:arma}
In this example we consider s-vine copula processes sharing the kpacf
of the ARMA(1,1) model with autoregressive parameter 0.95 and
moving-average parameter -0.85. Fixing $n=201$ we obtain Figure~\ref{fig:2}. Convergence appears to be fastest for the Gaussian and
Frank copula filters and slowest for the Clayton filter, followed by
the Joe filter; the Gumbel filter is an intermediate case
\end{example}

\begin{figure}[h]
\centering
\includegraphics[width=\textwidth]{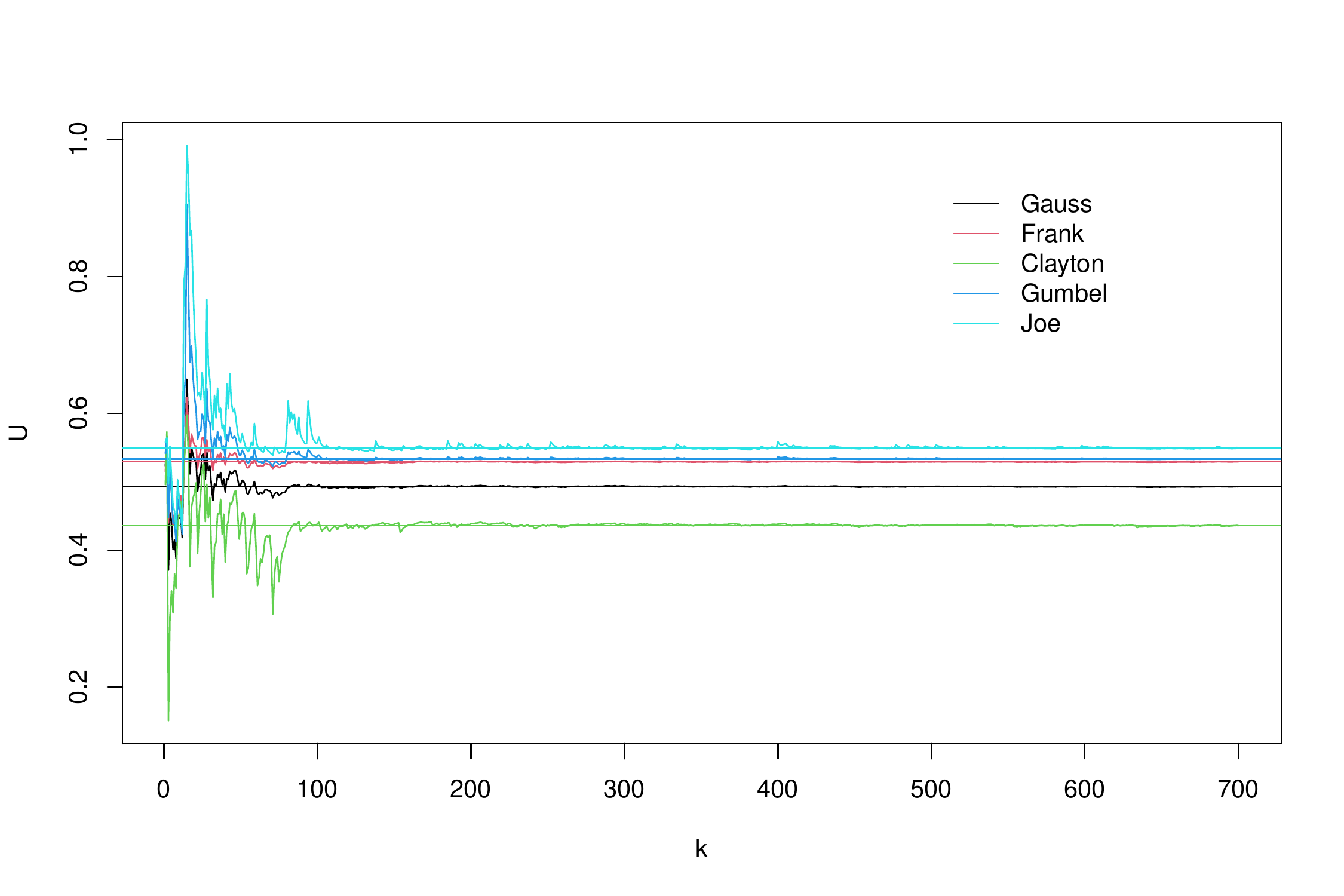}
\caption{Plots of $(k,
S_k(\bm{z}_{[n-k,n-1]},z_n))$ for $k \in
\{1,\ldots,n-1\}$ for the copula filters of ARFIMA(1,$d$,1)
models; see Example~\ref{example:arfima}. Horizontal lines show ultimate values $S_{n-1}(\bm{z}_{[1,n-1]},z_n)$.}
\label{fig:3}
\end{figure}

\begin{example}[Non-Gaussian ARFIMA(1,$d$,1) models]\label{example:arfima}
In this example we consider s-vine copula processes sharing the kpacf
of the ARFIMA(1,$d$,1) model with autoregressive parameter 0.95,
moving-average parameter -0.85 and fractional differencing
parameter $d = 0.02$. The latter implies that the pacf of the Gaussian process satisfies
$
  |\alpha_k| \sim 0.02/k$ as $k \to
  \infty$~\citep{bib:inoue-02}. The lack of absolute summability means
  that the Gaussian copula process does not satisfy the conditions of
  Theorem~\ref{theorem:causal-invertible}. It is an unresolved
  question as to whether any of these processes is causal.
Fixing $n=701$ we obtain
  Figure~\ref{fig:3}. For the realized series of innovations used in
  the picture, convergence appears to take place, but is extremely
  slow. The tail-dependent Clayton and Joe copulas appear to take
  longest to settle down.
\end{example}

An obvious practical solution that circumvents the issue of whether
the infinite-order process has a convergent causal representation is to truncate the copula
sequence $(C_k)_{k\in\N}$ so that $C_k = C^\perp$  for $k > p$ for
some relatively large but fixed value $p$. This places us back in the
setting of ergodic Markov chains but, by parameterizing models through the
kpacf, we preserve the advantages of parsimony.

\subsection{An example with real data}\label{sec:an-example-with}

\begin{figure}[h!]
\centering
\includegraphics[width=16cm,height=20cm]{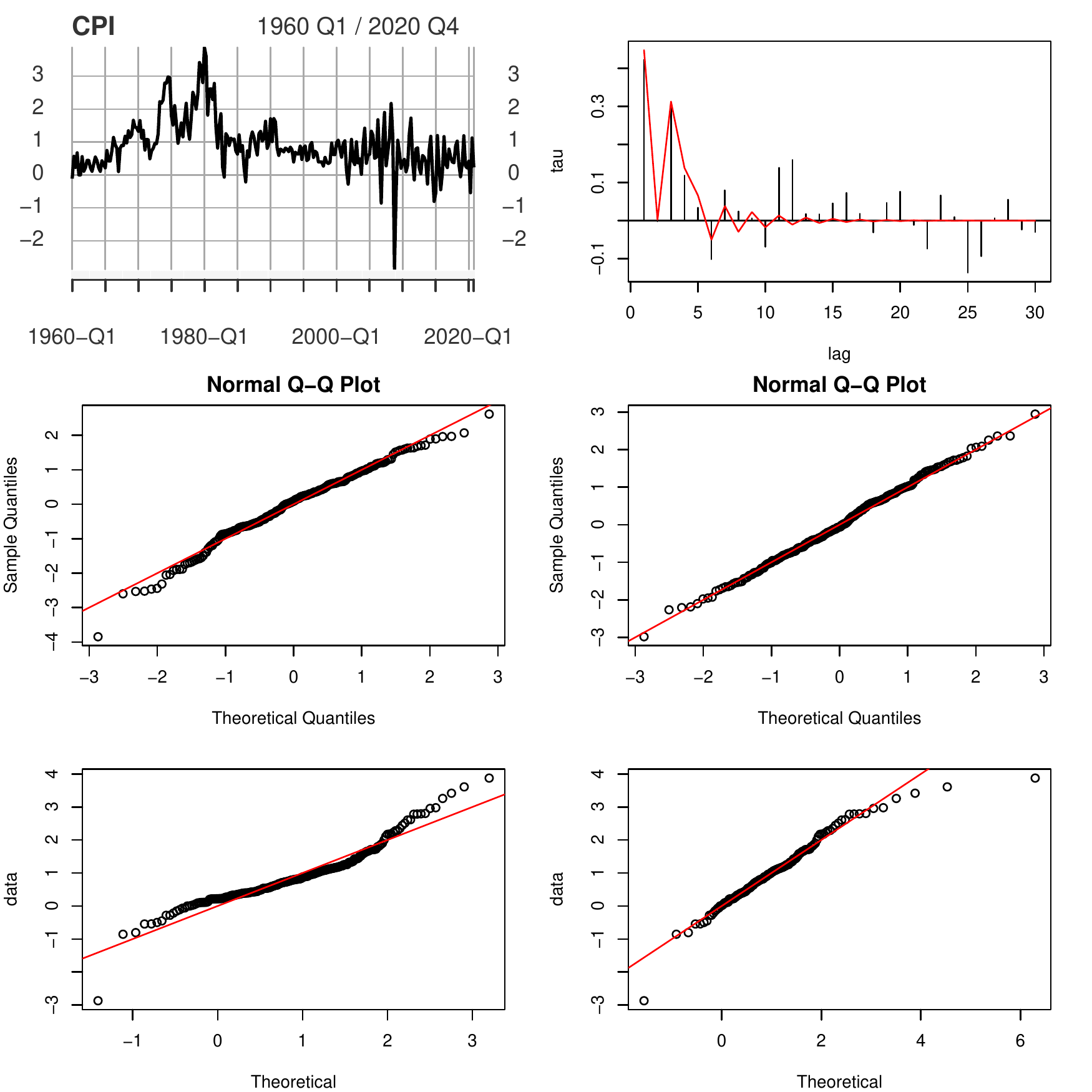}
\caption{Top row: log-differenced CPI data and estimated kpacf of
  s-vine copula process using Gumbel copula sequence. Middle row:
  QQ-plots for residuals from models based on Gaussian (left) and
  Gumbel (right)
  copula sequences. Bottom row: QQ-plots of the data against fitted
  normal (left) and skewed Student (right) marginal distributions.}
\label{fig:4}
\end{figure}

For this example we have used data on the US CPI (consumer
price index) taken from the OECD webpage. We analyse the
log-differenced time series of quarterly CPI values from the first
quarter of 1960 to the 4th quarter of 2020. The data are shown in the upper-left
panel of Figure~\ref{fig:4}; there are $n=244$ observations.

To establish a baseline model we use an automatic ARMA selection
algorithm and this selects an ARMA(5,1) model. We first address the
issue of whether the implied Gaussian copula sequence in an ARMA(5,1)
model can be replaced by Gumbel, Clayton, Frank or Joe
copula sequences (or 180 degree rotations thereof);
for any lag $k$ at which the estimated kpacf $\tau_k$ is negative we
retain a Gaussian copula and so the non-Gaussian copula sequences are actually
hybrid sequences with some Gaussian terms. The data $(x_1,\ldots,x_n)$ are
transformed to pseudo-observations
$(u_1,\ldots,u_n)$ on the
copula scale using the empirical distribution function and the s-vine copula process is estimated by
maximum-likelihood; this is the commonly used pseudo-maximum-likelihood
method developed for general copula inference
by~\cite{bib:genest-ghoudi-rivest-95} and adapted to time series by~\cite{bib:chen-fan-06b}. 

The
best model results from replacing Gaussian copulas with Gumbel
copulas and the improvement in the AIC is shown in the upper panel of Table~\ref{table2}; the
improvement in fit is strikingly large. While the presented results
relate to
infinite-order processes, we note that very similar result (not
tabulated) are obtained
by fitting s-vine copula processes of finite order where the kpacf is
truncated at lag 30. Parameter estimates for the infinite-order models
are given in Table~\ref{table1}.

The residual QQ-plots in the middle row of Figure~\ref{fig:4}
give further insight into the improved fit of the process with Gumbel copulas.
In the usual manner, residuals are reconstructions of the unobserved
innovation variables. If $(\widehat{R}_k)_{k\in\N}$ denotes the
sequence of estimated Rosenblatt forward functions, implied by the
sequence $(\widehat{C}_k)_{k\in\N}$ of estimated copulas, then
residuals $(z_1,\ldots,z_n)$ are constructed by setting $z_1 = u_1$ and
$z_t =\widehat{R}_{t-1}(\bm{u}_{[1,t-1]},u_t)$ for $t >1$. To
facilitate graphical analysis these are transformed onto the standard normal
scale so that the QQ-plots in the middle row of Figure~\ref{fig:4} relate to the values
$(\Phi^{-1}(z_1),\ldots,\Phi^{-1}(z_n)$ and are against a standard
normal reference distribution. The residuals from the
baseline Gaussian copula appear to deviate from normality whereas the
residuals from the Gumbel copula model are much better behaved; the latter pass a
Shapiro-Wilk test of normality (p-value = 0.97) whereas the former do
not (p-value = 0.01).

\begin{table}[ht]
\centering
\begingroup\setlength{\tabcolsep}{4pt}
\begin{tabular}{lrr}
  \toprule
 & No. pars & AIC \\ 
  \midrule
Gaussian copula process & 6 & -184.62 \\ 
  Gumbel copula process & 6 & -209.28 \\ 
  Gaussian process & 8 & 372.73 \\ 
  Gaussian copula process + skewed Student margin & 10 & 352.50 \\ 
  Gumbel copula process + skewed Student margin & 10 & 319.17 \\ 
   \bottomrule
\end{tabular}
\endgroup
\caption{Comparison of models by AIC: top panel relates to models for
  the pseudo-copula data $(u_1,\ldots,u_n)$ while  the lower panel
  relates to full models of the original data $(x_1,\ldots,x_n)$.} 
\label{table2}
\end{table}

\begin{table}[ht]
\centering
\begingroup\setlength{\tabcolsep}{4pt}
\begin{tabular}{rrrrr}
  \toprule
 & $\bm{\theta}^{\text{(Ga)}}$ & s.e. & $\bm{\theta}^{\text{(Gu)}}$ & s.e. \\ 
  \midrule
$\phi_1$ & -0.381 & 0.104 & -0.232 & 0.130 \\ 
  $\phi_2$ & 0.144 & 0.081 & 0.136 & 0.094 \\ 
  $\phi_3$ & 0.197 & 0.063 & 0.180 & 0.061 \\ 
  $\phi_4$ & 0.462 & 0.075 & 0.410 & 0.077 \\ 
  $\phi_5$ & 0.324 & 0.063 & 0.266 & 0.061 \\ 
  $\psi_1$ & 0.870 & 0.098 & 0.771 & 0.118 \\ 
   \bottomrule
\end{tabular}
\endgroup
\caption{Parameter estimates and standard errors for s-vine copula
  processes with Gaussian and Gumbel copula sequences fitted to the
  pseudo-copula data $(u_1,\ldots,u_n)$.} 
\label{table1}
\end{table}

The picture of the kpacf in the top right panel of Figure~\ref{fig:4}
requires further comment. This plot attempts to show how well the kpacf of
the fitted copula sequence matches the empirical Kendall partial
autocorrelations of the data. The continuous line is the kpacf of the Gumbel/Gaussian
copula sequence used in the best-fitting vine copula model of
$(u_1,\ldots,u_n)$.
The vertical bars show the empirical
Kendall partial autocorrelations of the data at each lag $k$. However, the method 
should really be considered as `semi-empirical' as it uses the fitted parametric copulas at lags $1,\ldots,k-1$
in order to construct the necessary data for lag $k$. The data used to
estimate an empirical lag $k$ rank correlation
are the
points
\begin{displaymath}
  \left\{ \Big(\widehat{\Rbackward}_{k-1}(\bm{u}_{[j-k+1,j-1]},u_{j-k}),
    \widehat{R}_{k-1}(\bm{u}_{[j-k+1,j-1]},u_{j}) \Big) ,\quad j =k+1,\ldots,n\right\},
\end{displaymath}
where $\widehat{R_k}$ and $\widehat{\Rbackward}_k$ denote the estimates of
forward and backward Rosenblatt functions;
it may be noted that these data are precisely the points at which the
copula density $c_k$ is evaluated when the model
likelihood based on $c_{(n)}$ in~(\ref{eq:4}) is
maximized.


We next consider composite models for the original data
$(x_1,\ldots,x_n)$ consisting of a marginal distribution and an s-vine
copula process. The baseline model is simply a Gaussian process with
Gaussian copula sequence and Gaussian marginal distribution. We
experimented with a number of alternatives to the normal marginal
and obtained good results with the skewed Student
distribution from the family of skewed distributions proposed
by~\cite{bib:fernandez-steel-98}. Table~\ref{table2} contains results
for models which combine the Gaussian and Gumbel copula sequences with
the skewed Student margin; the improvement obtained by using a Gumbel sequence
with a skewed Student margin is clear from the AIC values.
The
QQ-plots of the data against the fitted marginal distributions in the bottom row of
Figure~\ref{fig:4} also show the superiority of the skewed Student
to the Gaussian distribution for this dataset.

The fitting
method used for the composite model results in Table~\ref{table2} is the two-stage IFM (inference functions for
margins) method of~\cite{bib:joe-97} in which the margin is estimated
first, the data are transformed to approximately uniform using the
marginal model, and the copula process is estimated by ML in a second step.

\section{Conclusion}\label{sec:conclusion}

The s-vine processes provides a class of tractable stationary
models that can capture non-linear and non-Gaussian serial dependence behaviour
as well as any continuous marginal behaviour. By defining models of
infinite order and using the approach based on the Kendall partial
autocorrelation function (kpacf), we obtain a very natural generalization of
classical Gaussian processes, such as Gaussian ARMA or ARFIMA.

The models are straightforward to apply. The parsimonious
parametrization based on the  kpacf makes maximum likelihood inference
feasible. Analogues of many of the standard tools for time series
analysis in the time domain are available, including estimation methods
for the kpacf and residual plots that shed light on the
quality of the fit of the copula model. By separating the issues of
serial dependence and marginal modelling, we can obtain
bespoke descriptions of both aspects that avoid the compromises
of the more `off-the-shelf' classical approach. The example of
Section~\ref{sec:an-example-with} indicates the kind of gains that can
be obtained; it seems likely that many empirical applications of
classical ARMA could be substantially improved by the use of models in
the general s-vine class. In combination with
v-transforms~\citep{bib:mcneil-20} s-vine models could also be used to
model data showing stochastic volatility following the approach in~\citet{bib:bladt-mcneil-21}.

The approach we have adopted should also be of interest to theoreticians
as there are a number of challenging open questions to be
addressed. While we have proposed definitions of causality and
invertibility for general s-vine processes, we currently lack a mathematical
methodology for checking convergence of causal and invertible
representations for sequences of non-Gaussian pair copulas.

There are some very interesting questions to address
about the relationship between the partial copula sequence
$(C_k)_{k\in\N}$, the rate of convergence of causal representations
and the rate of ergodic mixing of the resulting
processes. The example of Figure~\ref{fig:1} indicates that, even
for a finite-order process,
some very extreme models can be constructed that mix extremely slowly.
Moreover, Example~\ref{example:arfima} suggests that non-Gaussian copula sequences serve to
further elongate memory in long-memory processes and this raises
questions about the effect of the tail dependence properties of the
copula sequence on rates of convergence and length of memory.

It would also be of interest to confirm our
conjecture that the pragmatic approach adopted in
Section~\ref{sec:pract-appr-non}, in which the kpacf of the (infinite) partial
copula sequence $(C_k)_{k\in\N}$ is matched to that of a stationary and ergodic
Gaussian process, always yields a stationary and ergodic s-vine model,
regardless of the choice of copula sequence. However, for practical
applications, the problem can be obviated by truncating the copula sequence at some large
finite lag $k$, so that we
are dealing with an ergodic Markov chain as in Section~\ref{sec:finite-order}.

\appendix
\section{Proofs}\label{sec:proofs}
 \subsection{Proof of Proposition~\ref{prop:reversibility}}

      In this proof we use the notation $(\bm{u})_i$ to denote the $i$th
      component of a vector $\bm{u}$ and $\bm{u}_{-i}$ to denote the
      vector $\bm{u}$ with $i$th component removed. 
An exchangeable copula satisfies $C_k(u,v) = C_k(v,u)$ for all
$u,v$ and hence $h_k^{(2)}(u,v) =
h_k^{(1)}(v,u)$. From this it follows that $\Rbackward_1(u,x) = h_1^{(2)}(x,u) =
h_1^{(1)}(u,x) = \Rforwardb_1(u,x)$. Part (i) follows by induction
using the facts that for $\bm{u} =(u_1,\ldots,u_k)^\top \in (0,1)^k$ we
have $(\overline{\bm{u}})_k = u_1$, $\overline{\bm{u}}_{-k} = (u_{k},\ldots,u_2)^\top$ and
$\overline{\overline{\bm{u}}_{-k}} = \bm{u}_{-1}$. We have that
\begin{align*}
  \Rbackward_k(\overline{\bm{u}}, x) &= h_k^{(2)}\Big(\Rbackward_{k-1}(\overline{\bm{u}}_{-k}, x),
  \Rforwardb_{k-1}(\overline{\bm{u}}_{-k},(\overline{\bm{u}})_k)\Big) \\
  &= h_k^{(1)}\Big(\Rforwardb_{k-1}(\overline{\bm{u}}_{-k}, u_1),
    \Rbackward_{k-1}(\overline{\bm{u}}_{-k},x)\Big)
  = h_k^{(1)}\Big(\Rbackward_{k-1}(\bm{u}_{-1},u_1),
    \Rforwardb_{k-1}(\bm{u}_{-1},x)\Big) = \Rforwardb_k(\bm{u}, x).
\end{align*}
For part (ii) we observe that for any $k >1$ the implication of
part(i) is that for any $t,s\in\N$ the conditional
distribution of $U_{t+k} \mid U_{t+1}=u_1,\ldots,U_{t+k-1} = u_{k-1}$ is the
same as that of $U_{s+1} \mid U_{s+2} =u_{k-1}, \ldots, U_{s+k} = u_1$. It
easily follows
that $(U_{t+1},\ldots,U_{t+k}) \eqdis (U_{s+k},\ldots,U_{s+1})$ which
proves reversibility of the process.
\subsection{Proof of Proposition~\ref{prop:gaussian-processes}}
  If $(X_t)$ is
a Gaussian process its marginal distributions of all orders are 
multivariate Gaussian. The general d-vine copula decomposition
in~(\ref{D_vine_density}) can be applied to each $n$-dimensional
marginal density. Since the conditional distributions of pairs $(X_{j-k}, X_{j})$
given intermediate variables are 
bivariate Gaussian distributions with covariance
matrices that do not depend on the conditioning variables, the
simplifying assumption holds
for each pair copula density in~(\ref{D_vine_density})~\citep[see
also][pages 106--108]{bib:joe-15}. The
stationarity assumption ensures that the joint density of the
$n$-dimensional copula takes the form~(\ref{eq:4}).

Conversely, an
s-vine process with Gaussian marginal density and Gaussian pair
copulas is a stationary process with $n$-dimensional marginal
densities of the form~\eqref{eq:1}. These are the
densities of multivariate
Gaussian distributions and
the resulting process is a Gaussian process.

\subsection{Proof of Proposition~\ref{prop:gaussian-case}}
                      Let $(Z_k)_{k\in\N}$ be a sequence of iid
                      standard uniform variables and $(U_k)_{k\in\N}$ a sequence of uniform
                      random variables generated by setting $U_1 =
                      Z_1$ and $U_k =
                      R^{-1}_{k-1}\Big((U_{1},\ldots,U_{k-1})^\top,Z_k\Big)$
                      for $k>1$ where $(R_k)_{k\in\N}$ denotes the sequence of
                      Rosenblatt functions associated with the
                      sequence of Gaussian pair copulas
                      $(C_k)_{k\in\N}$. Moreover, let $(X_k)_{k\in\N}$ be a
                      sequence of standard Gaussian variables defined
                      by setting $X_k = \Phi^{-1}(U_k)$ for all $k$.
                      
                      It follows that, for any $k\geq 1$,  $(X_1,\ldots,X_{k+1}) \sim N_{k+1}(\bm{0},
P_{k+1})$ where $P_{k+1}$ is the $(k+1)$-dimensional correlation
matrix implied by the acf $(\rho_i)_{i\in\N}$ of $(X_i)_{i\in\N}$ as in~(\ref{eq:22}). The standard result for the conditional distribution of a
multivariate normal implies that
\begin{displaymath}
  X_{k+1} \mid X_1 = x_1, \ldots,
  X_k = x_k \sim N\Big( \overline{\bm{\rho}}_k^\top
  P_k^{-1}  (x_1,\ldots,x_k)^\top,\; 1 -
  \bm{\rho}_k^\top P_k^{-1} \bm{\rho}_k\Big),
\end{displaymath}
where $\bm{\rho}_k=(\rho_1,\ldots,\rho_k)^\top$ as in~(\ref{eq:22})
and $\overline{\bm{\rho}}_k$ is the reversed vector. The mean of the
conditional distribution is the best linear predictor of $X_{k+1}$ and
the variance of the conditional distribution is the mean squared
prediction error; let us write the former as $\sum_{j=1}^k
\phi^{(k)}_j x_{k+1-j}$, where $\phi^{(k)}_{j} = (P_k^{-1}  \bm{\rho}_k)_j$,  and the latter as $\sigma_k^2$. We then have
\begin{eqnarray*}
  R_k(\bm{u}, x )  & = & \P( U_{k+1} \leq x \mid U_1 =
                         u_1,\ldots,U_{k} = u_{k})\\
  & = & \P( X_{k+1} \leq \Phi^{-1}(x) \mid X_1 =
                         \Phi^{-1}(u_1),\ldots,X_{k} = \Phi^{-1}(u_{k}))\\
                   & = & \Phi\left(
                         \frac{\Phi^{-1}(x) - \sum_{j=1}^k
                        \phi^{(k)}_{j}
                        \Phi^{-1}(u_{k+1-j})}{\sigma_k}
                         \right).
\end{eqnarray*}
The expression $\sigma_k^2 = \prod_{j=1}^k (1-\alpha_j^2)$ and the
recursive formula~(\ref{eq:11}) for the coefficients $\phi_j^{(k)}$
follow from the Durbin--Levinson Algorithm;
see~\citet{bib:brockwell-davis-91}, Proposition 5.2.1.

It follows from~(\ref{eq:9}) that, for $k >0$,
                       \begin{equation}\label{eq:12}
        U_{k+1} = R_{k}^{-1}((U_{1},\ldots,U_{k})^\top,
                       Z_{k+1}) =     \Phi\left(
                         \sigma_{k}\Phi^{-1}(Z_{k+1}) + \sum_{j=1}^{k}
                        \phi^{(k)}_{j}
                        \Phi^{-1}(U_{k+1-j})
                         \right)            
                       \end{equation}
 which may be written in terms of the variables  $(X_k)_{k\in\N}$ as
                         $X_{k+1} = \sigma_{k} \epsilon_{k+1} + \sum_{j=1}^{k}
                        \phi^{(k)}_{j}
                        X_{k+1-j}$
where we introduce the further notation
$\epsilon_{k} = \Phi^{-1}(Z_k)$ for all $k\in\N$. An inductive argument then shows that
this may be written in the form
$X_{k+1} = \sigma_{k} \epsilon_{k+1} +\sum_{j=1}^{k}
                        \psi^{(k)}_{j}
                        \epsilon_{k+1-j} $
                      with the coefficients $\psi_j^{(k)}$ as defined
                      in~(\ref{eq:17b}). Equation~(\ref{eq:17}) then
                      follows easily from~(\ref{eq:15}).

\subsection{Proof of Theorem~\ref{theorem:causal-invertible}}
As in the proof of Proposition~\ref{prop:gaussian-case} we introduce the notation
$(X_t)_{t\in\Z}$ and $(\epsilon_t)_{t\in\Z}$ where $X_t =
\Phi^{-1}(U_t)$ and $\epsilon_t = \Phi^{-1}(Z_t)$. For fixed $k$ the formulas $U_t =
       S_k((Z_{t-k},\ldots,Z_{t-1})^\top,Z_t)$ and $Z_t = R_k((U_{t-k},\ldots,U_{t-1})^\top,U_t)$ translate to
        \begin{displaymath}
                               X_{t} = \sum_{j=0}^{k}
                        \psi^{(k)}_{j}
                        \epsilon_{t-j}\quad\text{and}\quad  \epsilon_{t} = \sum_{j=0}^{k}
                        \tilde{\phi}^{(k)}_{j}
                        X_{t-j}
        \end{displaymath}
    where $\tilde{\phi}^{(k)}_0 = 1/\sigma_k$ and
    $\tilde{\phi}^{(k)}_j = - \phi^{(k)}_j/\sigma_k$ for $j >1$.

    \citet[Theorem 6]{bib:debowski-07} shows that under
    Assumption~\ref{assumption:partialcorrelationsum} the limiting
    representations
       \begin{equation}\label{eq:19}
                               X_{t} = \sum_{j=0}^{\infty}
                        \psi_{j}
                        \epsilon_{t-j}\quad\text{and}\quad \epsilon_{t} = \sum_{j=0}^{\infty}
                        \tilde{\phi}_{j}
                        X_{t-j}
\end{equation}
hold where $(\tilde{\phi}_j)_{j\in\Z_+}$ and $(\psi_j)_{j\in\Z_+}$ are
sequences such that
    $\lim_{k\to\infty} \tilde{\phi}_j^{(k)} = \tilde{\phi}_j$ and
    $\lim_{k\to\infty} \psi_j^{(k)} = \psi_j$ and such that $\psi_j = \cov(X_t,Z_{t-j})$, $\tilde{\phi}_0 =
    1/\psi_0$ and $\sum_{j=0}^k \tilde{\phi}_j \psi_{k-j} = 0$ for $k
    >0$; the series in the rhs of~\eqref{eq:19} converge absolutely, almost surely. Moreover, under Assumption~\ref{assumption:partialcorrelationsum} we also
    have that the terms $\sigma_k ^2 = \prod_{j=1}^k(1-\alpha_j^2)$ 
    converge to a finite limit $\sigma$ and so we can introduce
    a sequence $(\phi_j)_{j\in\N}$ such that
    $\phi_j =-\sigma \tilde{\phi}_j$ and write
    \begin{displaymath}
      X_{t} = \sigma \epsilon_t + \sum_{j=1}^{\infty}
                        \psi_{j}
                        \epsilon_{t-j}\quad\quad\text{and}\quad\quad
                        \epsilon_{t} = \sigma^{-1} X_t - \sigma^{-1}\sum_{j=1}^{\infty}
                        \phi_{j}
                        X_{t-j}.
                      \end{displaymath}
Finally, by Proposition \ref{prop:gaussian-case}, equations~(\ref{eq:17}) and~(\ref{eq:18}) are seen to be a restatement of the latter formulas in
terms of the Rosenblatt functions.

\section{Markov chain analysis}\label{appendix:mark-chain-analys}

The Markov chain
specified by~(\ref{eq:3}) under
Assumption~\ref{assumption:smooth-copulas} is a well-behaved example
of a chain on a general state space. The properties of the process can
be verified by standard arguments which are collected here for completeness.

\paragraph{Invariance.} The transition kernel of the Markov chain is given by
\begin{align*}
  \mathsf{P}(\bm{u}, A) &=  \P\left(\bm{U}_2 \in A \mid \bm{U}_1=\bm{u} \right) =
                 \P\left(F(\bm{U}_1,Z) \in A  \mid \bm{U}_1=\bm{u}\right) = \prod_{i=2}^p \indicator{u_i \in
                    A_{i-1}} \int_{A_p} f_p(\bm{u}, x) \rd x
\end{align*}
for a set $A = A_1\times \cdots \times A_p \subseteq \mathcal{X}$. Writing $\pi$ for
the probability measure implied by $C_{(p)}$ and using~(\ref{eq:1}), we have that
\begin{align*}
  \int_{(0,1)^p} \pi(\rd \bm{u}) \mathsf{P}(\bm{u}, A) &=
    \int_{(0,1)^p} c_{(p)}(u_1,\ldots,u_p)        \prod_{i=2}^p \indicator{u_i \in
                    A_{i-1}} \left(\int_{A_p} f_p(\bm{u}, x) \rd x \right)\rd \bm{u}
  \\
  &= \int_0^1\int_{A_1} \cdots \int_{A_{p-1}}  c_{(p)}(u_1,\ldots,u_p)
    \left(\int_{A_p}f_p(\bm{u}, x) \rd x\right)\rd u_1 \cdots \rd u_p \\
  &= \int_0^1 \int_{A_1} \cdots \int_{A_p}
    c_{(p+1)}(u_1,\ldots,u_p,u_{p+1}) \rd u_1 \cdots \rd u_{p+1} \\
  &= \int_{A_1} \cdots \int_{A_p}
    c_{(p)}(u_2,\ldots,u_p,u_{p+1}) \rd u_2 \cdots \rd u_{p+1} = \pi(A)
\end{align*}
showing that $\pi$ is an invariant measure.

\paragraph{Irreducibility.} 
A process is $\phi$-irreducible if there is
a measure $\phi$ on $\mathcal{X}$ such that for every
set $A \subseteq \mathcal{X}$ with $\phi(A) >0 $ and every $\bm{u} \in (0,1)^p$
there exists $n =n(\bm{u},A) >0$ such that $\mathsf{P}^n(\bm{u},A) >0$. In our case it
suffices to take $n=p$, independent of $\bm{u}$ and $A$, and $\phi$ to
be Lebesgue measure.
After $p$-fold iteration of the Markov updating scheme in~(\ref{eq:3})
we obtain the random vector $\bm{U}_{p+1} =
(U_{p+1},\ldots,U_{2p})^\top$ and for a set $A = A_1 \times \cdots
\times A_p$ with positive Lebesgue measure we have
\begin{align*}
  \mathsf{P}^p(\bm{u}, A) &= \P(\bm{U}_{p+1} \in A \mid \bm{U}_1 = \bm{u}) \\
&= \P(U_{p+1} \in A_1, \ldots, U_{2p} \in A_p \mid U_1
  = u_1, \ldots, U_p = u_p) \\
  &= \int_{A_1} \cdots \int_{A_p} f_{U_{p+1},\ldots,U_{2p} \mid U_1,
    \ldots, U_p}(x_{1}, \ldots, x_{p} \mid u_1, \ldots, u_p) \rd
    x_{1} \cdots \rd x_{p}
\end{align*}
and this probability is strictly positive if the integrand is strictly
positive. Since the function in the integrand satisfies
\begin{align*}
 f_{U_{p+1},\ldots,U_{2p} \mid U_1,
    \ldots, U_p}(u_{p+1}, \ldots, u_{2p} \mid u_1, \ldots, u_p) 
  &=
  \prod_{j=1}^p f_p((u_j,\ldots,u_{j+p-1})^\top,u_{j+p})
\end{align*}
and
the conditional densities $f_p(\bm{u},x)$ as defined in~(\ref{eq:1}) are products of strictly
positive pair-copula densities, the result follows.

\paragraph{Recurrence.}
Since the Markov chain is $\phi$-irreducible and admits an invariant
probability measure, it is a positive recurrent chain. The absolute
continuity of the transition kernel with respect to Lebesgue measure
(exploited above)
also means it is 
a Harris recurrent chain: for any point $\bm{x} \in \mathcal{X}$ and
any set $A$ with invariant measure $\pi(A) =1$, either $\bm{x} \in A$
or, if not, $\mathsf{P}(\bm{x},A) =1$ so that it is certain that the time to
entering $A$ is finite and this is a condition for Harris recurrence;
see, for example,~\citet[Theorem 6(v)]{bib:roberts-rosenthal-06}.

\paragraph{Aperiodicity.}
A periodic chain would cycle through $m \geq 2$ disjoint subsets of
the state space $\mathcal{X}_1, \ldots,\mathcal{X}_m$, each satisfying
$\phi(\mathcal{X}_i) >0$, in successive steps. If such behaviour were to occur we could find an $n\geq p$ such
that $\mathsf{P}^n(\bm{u},\mathcal{X}_2) = \P(\bm{U}_{n+1} \in \mathcal{X}_2 \mid \bm{U}_1 =\bm{u}) = 1$
for all $\bm{u} \in \mathcal{X}_1$, which would imply that
$\mathsf{P}^n(\bm{u},\mathcal{X}_1) = 0$ for all
$\bm{u} \in \mathcal{X}_1$. However, since $\phi(\mathcal{X}_1) >0 $,
the argument used to establish the $\phi$-irreducibility of the
process can be repeated to show that $\mathsf{P}^n(\bm{u},\mathcal{X}_1) > 0$ for
all $\bm{u} \in \mathcal{X}$,
which yields a contradiction.

\bibliographystyle{apalike}
\setcitestyle{authoryear,open={(},close={)}}
\newcommand{\noopsort}[1]{}

\end{document}

%% file: preamble.tex
\usepackage{geometry,setspace}
\small\normalsize

\usepackage{hyperref}
\hypersetup{
    colorlinks=true,
    citecolor=teal,
    linkcolor=teal
}
\usepackage{caption}
\usepackage{enumerate}
\usepackage{graphicx}
\usepackage{rotating,multirow}
\usepackage[table]{xcolor}
 \usepackage{amsmath,amsfonts,amssymb,mathrsfs,amsthm}
 \usepackage{mathtools}
\usepackage{bbm}
\usepackage{bm}
\usepackage[authoryear, round]{natbib}
\usepackage{booktabs}
\usepackage{pdflscape}
\usepackage{todonotes}
\usepackage{authblk}
\usepackage{xr}
\usepackage{tikz-cd}
\usepackage{framed}

\usepackage{xfrac}

\usepackage{caption}
\usepackage[T1]{fontenc}
\input{glyphtounicode}
\newtheorem{proposition}{Proposition}
\newtheorem{theorem}{Theorem}

\theoremstyle{definition}
\newtheorem{definition}{Definition}
 \newtheorem{assumption}{Assumption}
\newtheorem{remark}{Remark}
\newtheorem{example}{Example}

\newcommand{\indicator}[1]{\ensuremath{I_{\{#1\}}}}

\usepackage{accents}

\newcommand{\Rforward}{\ensuremath{R^{(1)}}}
\newcommand{\Rforwardb}{\ensuremath{R}}
\newcommand{\Rbackward}{\ensuremath{R^{(2)}}}

\renewcommand{\P}{\mathbb{P}}
\newcommand{\E}{\mathbb{E}}
\newcommand{\R}{\mathbb{R}}

\newcommand{\Z}{\mathbb{Z}}
\newcommand{\N}{\mathbb{N}}

\newcommand{\eqdis}{\stackrel{\text{\tiny d}}{=}}

\newcommand{\cov}{\operatorname{cov}}

\renewcommand{\leq}{\leqslant}
\renewcommand{\geq}{\geqslant}

\newcommand{\rd}{\mathrm{d}}